\documentstyle[preprint,aps]{revtex}
\begin{document}

%
% Title Page
%

\title{{\it Ab initio} molecular dynamics using density based energy
functionals: application to ground state geometries of some small
clusters.}
\author{Dinesh Nehete, Vaishali Shah and D. G. Kanhere \cite{email}}
\address{Department of Physics, University of Poona, Pune 411 007, India}
%\date{\today}
\maketitle

\begin{abstract}

The ground state geometries of some small clusters have been obtained
via ab initio molecular dynamical simulations by employing density
based energy functionals. The approximate kinetic energy functionals
that have been employed are the standard Thomas-Fermi $(T_{TF})$ along
with the Weizsacker correction $T_W$ and a combination
$F(N_e)T_{TF}  +  T_W$. It is shown that the functional involving $F(N_e)$
gives superior charge densities and bondlengths over the standard
functional. Apart from dimers and trimers of Na, Mg, Al, Li, Si,
equilibrium geometries for $Li_nAl, n=1,8$ and $Al_{13}$ clusters
have also been reported. For all the clusters investigated, the method
yields the ground state geometries with the correct symmetries with
bondlengths within 5\% when compared with the corresponding results
obtained via full orbital based Kohn-Sham method. The method is fast
and a promising one to study the ground state geometries of large clusters.

\noindent PACS Numbers : 71.10, 31.20G, 02.70N, 36.40

\end{abstract}

\newpage

%
% Introduction
%

\section {Introduction}

During the last few years the technique of first principles molecular dynamics
(MD), initiated by Car and Parrinello (CP) \cite{car,rem}, has emerged as a
powerful tool for investigations of structural, electronic and thermodynamic
properties of large scale systems. The standard implementation of this method
which is based on density functional theory is via Kohn-Sham (KS) orbitals.
Such orbital based algorithms scale as  $N_a^3$, $N_a$ being the
number of atoms in the system.
Quite clearly, such methods turn out to be computationally expensive
for system sizes over about 100 atoms \cite{footnote}.
Recently, approaches  based on total energy functionals, which depend on
charge density only or orbital free density functionals have been proposed
\cite{pear,shah,govind}.
These methods are based on approximate representation of kinetic energy (KE)
functionals and offer an attractive alternative for investigating large
scale systems. Since the method is orbital free i.e there are no wavefunctions
to handle, there is no computationally expensive orthogonality constraint and
the methods scale linearly with system size. In addition, these methods are
shown to yield stable dynamics even with large timesteps, a highly
desirable feature for molecular dynamics simulations.

It is clear that the utility of these methods is critically dependent on
their ability to investigate the systems of interest with acceptable
accuracy, at least for a class of physical properties.
Madden and coworkers have investigated structural and thermodynamic
properties of some simple metals with considerable success.
For example, the dynamic structure factor of liquid Sodium and static structure
factor, vacancy formation energy, free energies of point defects as well as
phonon dispersion curves of Sodium \cite{pear,smar} are well described by this
method. The method has also been applied for ground state configurations
of c-Si and H/Si (1 0 0) surface \cite{govind} and for geometries
of some silicon clusters \cite{niranjan} and a good agreement has been found
with experiments as well as with other calculations. However majority of
the calculations reported so far have been performed on extended systems.

In the present work, we focus our attention on studying the ground state
and energetically low lying structures of clusters, a field of current
interest. Obviously, due to the approximate nature of the KE functionals
the bondlengths and binding energies will not be obtained with the same
level of accuracy as the KS orbital based methods. However, it is of
considerable interest to examine whether such a method is capable of
yielding the correct shapes (i.e the right symmetries) of clusters
by employing Car-Parrinello simulated annealing methods.
If desired the KS method can then be used to search the local minimum
around structures obtained by Orbital Free Method (OFM) in `quenching' mode.
This can be a computationally tractable way to avoid the long and costly
simulated annealing runs of the orbital based KS molecular dynamics.

Towards this end we have carried out a number of calculations on a
variety of representative small clusters of simple metals.
Specifically, we have investigated dimers and trimers of Na, Mg, Al, Li, Si,
small clusters of $Na_n, (n=6,8)$, $Li_nAl, (n=1,8)$ and $Al_{13}$.
These systems are representative of the small metal atom clusters of
current interest and more accurate KS based results have been reported.
Hence, it is possible to make an assessment of the present method by
comparing the bondlengths and geometries with the reported ones.

The question of appropriate choice of kinetic energy functionals has been
addressed by Smargiassi and Madden \cite{smarg}.
They have investigated a family of kinetic energy functionals which incorporate
exact linear response properties. All such KE functionals are based on the
Thomas Fermi (TF) and the Weizsacker correction term. Since our interest is
in finite size systems, we have chosen to use simple KE functionals. These
functionals have been previously used in the study of atoms and molecules.
However, it must be mentioned that significant progress has been made towards
improving the KE functionals notably by DePristo and Kress, Wang and Teter
\cite{depristo,wang}.

In the next section we briefly discuss the method, the KE functionals used and
give the relevant numerical details. This is then followed by the results and
discussion.

\section {Formalism and Computational details}

\subsection { Total Energy Calculation}

The total energy of a system of $N_e$ interacting electrons and $N_a$ atoms,
according to the Hohenberg-Kohn theorem \cite{hoh,ksh}, can be uniquely
expressed as a functional of the electron density $\rho({\bf r})$ under an
external potential due to the nuclear charges at coordinates ${\bf R}_n$,
\begin{equation}
   E\Bigl[\rho,\{{\bf R}_n\}\Bigr] = T[\rho]
                                  + E_{xc}[\rho]
                                  + E_c[\rho]
                                  + E_{ext}\Bigl[\rho,\{{\bf R}_n\}\Bigr]
                                  + E_{ii}\Bigl(\{{\bf R}_n\}\Bigr),
\end{equation}
where
$E_{xc}$ is the exchange-correlation energy,
$E_c$ is the electron-electron Coulomb interaction energy.
The electron-ion interaction energy $E_{ext}$ is given by
\begin{equation}
   E_{ext}\Bigl[\rho,\{{\bf R}_n\}\Bigr] =  \int{V({\bf r})
                                                 \rho({\bf r}) d^3r}
\end{equation}
where $V({\bf r})$ is the external potential, usually taken to be a convenient
pseudopotential \cite{bach}.
The last term in Eq. (1), $E_{ii}$, denotes the ion-ion interaction energy.
The first term in Eq. (1), the KE functional, is usually approximated as
\begin{equation}
   T[\rho] =  T_{TF}[\rho] + T_W[\rho]
\end{equation}
where $T_{TF}[\rho]$ is the Thomas-Fermi term, exact in
the limit of homogeneous density, and has the form
\begin{equation}
   T_{TF}[\rho] =   \frac{3}{10}(3\pi^2)^{\frac{2}{3}}
               \int{\rho({\bf r})^{5/3} d^3r}
\end{equation}
and $T_W[\rho]$ is the gradient correction due to Weizsacker, given as
\begin{equation}
   T_W[\rho]   =    \frac{\lambda}{8} \int{\frac{\nabla\rho({\bf r})
               \cdot\nabla\rho({\bf r}) d^3r}{\rho({\bf r})}}
\end{equation}
which is believed to be the correct asymptotic behavior of T$[\rho]$ for
rapidly varying densities.
Instead of $\lambda = 1$, the original Weizsacker value,
$\lambda = \frac{1}{9}$ and $\lambda = \frac{1}{5}$ \cite{parr} are
also commonly used. It has been argued that for rapidly varying densities,
which is the case for finite size clusters a more appropriate kinetic
energy functional would be a following combination of these two
terms\cite{acharya}
\begin{equation}
   T[\rho] =  F(N_e) T_{TF}[\rho] + T_W[\rho]
\end{equation}
where the factor $F(N_e)$ \cite{gaz} is
\begin{equation}
   F(N_e)   = \bigglb(1-\frac{2}{N_e}\biggrb)
            \bigglb(1-\frac{A_1}{N_e^{\frac{1}{3}}}
           +\frac{A_2}{N_e^{\frac{2}{3}}}\biggrb)
\end{equation}
with optimized parameter values $A_1 = 1.314$ and $A_2 = 0.0021$ \cite{ghosh}.
This functional which includes the full contribution of the Weizsacker
correction describes the response properties of the electron gas well. This
functional has been used for investigating atoms and molecules with reasonable
success.

We briefly describe our procedure, details of which can be found in
\cite{shah}. The total energy of the system (Eq. (1)), is minimized for fixed
ionic positions using the conjugate gradient method \cite{press} which forms
the starting point for molecular dynamics.
The trajectories of ions and the fictitious electron dynamics are then
simulated using Lagrange's equations of motion which are solved by
Verlet algorithm \cite{car}.
The stability of CP dynamics has been discussed in \cite{pear} in the context
of density based methods and timesteps of the order to 50 a.u. have been
successfully used.
We have verified that by appropriate adjustment of the fictitious electron mass
the CP dynamics remains very stable for over 10000 iterations with a timestep
of 40 a.u. in the present calculations of clusters.
Typically, for free dynamics the grand total energy which is the
sum of the kinetic energy of ions, kinetic energy of electrons and the
potential
energy of the system remains constant to within $10^{-5}$ a.u.

For the calculations of the ground state structures for dimers and trimers of
Na, Mg, Al, Li and Si a periodically repeated unit cell
of length 26 a.u. with a 54 $\times$ 54 $\times$ 54 mesh and timestep $\Delta
t \sim$  10 to 20 a.u. was used.  For the rest of the small clusters the
calculations
were done on a unit cell of length of 30 a.u with a 54 $\times$ 54 $\times$ 54
mesh. We have chosen to use the plane wave expansion on the entire fast
fourier transform mesh without any truncation yielding the energy cutoff of
95 Rydberg.
It must be mentioned that, due to the orbital free calculations the number
of fast fourier transforms per iteration are constant irrespective of the
number of electrons in the system.
For clusters, the ground state configurations are obtained either by
starting with different initial configurations and then quenching the
structures or by dynamical simulated annealing where the cluster is heated
to $300-350^{\circ} K$ and then cooled very slowly.
In all the cases the stability of the final ground state configurations has
been tested by reheating the clusters and allowing them to span the
configuration space for a few thousand iterations and then cooling them
to get the low energy configurations.

\section {Results and Discussion}

In this section, we first discuss the results for the equilibrium
bondlengths and binding energies of dimers and trimers of Na, Mg, Al, Li,
Si along with their KS results.
All the results presented here are obtained
with energy convergence up to $10^{-13}$ for total energy minimization.

Table I shows the equilibrium bondlengths and binding energies
for dimer and trimer systems using different
kinetic energy functionals. These results have been compared with full
nonlocal pseudopotential KS calculations. A few representative results
using $\lambda = \frac{1}{5}$ have been given. It can be seen that for
$\lambda = \frac{1}{5}$ the trend is similar to the $\lambda = \frac{1}{9}$
functional and there is no significant improvement in the results.
Clearly, the results involving $F(N_e)$ functional show significant
improvement over $\lambda = \frac{1}{9}$  (with the exception of Mg)
and are in reasonable agreement with the bondlengths obtained by the
KS method.
The error in the bondlengths is around 10\%.
It is known that such methods based on approximate KE functionals are not
expected to give accurate binding energies. One notable feature of
the binding energy comparison is the considerable improvement by $F(N_e)$ over
$\lambda = \frac{1}{9}$ (excepting again the case of Mg). The results
for the Na, Li, Si trimer binding energies are not given because these
are Jahn-Teller distorted isoscales triangles and the present method
yields equilateral triangle geometries. Clearly, such density based
methods are unable to reach the Jahn-Teller distorted geometries.

The quality of the charge densities obtained by this method can be gauged
by comparing them with the KS charge density.
In Fig. 1 we have plotted the self-consistent charge densities obtained
using the functionals involving $F(N_e)$ (curve a) and
$\lambda = \frac{1}{9}$ (curve b) with the KS charge density (curve c)
for Al dimer along the axis joining the atoms. The ionic positions are marked
by arrows on the plot.
The KS charge density has been obtained using the identical pseudopotentials
and the same cell size as in the case of OFM. Three prominent features
can be observed.
\begin{enumerate}
\item Overall the $F(N_e)$ functional densities compare very well with the KS
densities except at the origin where both the $F(N_e)$
and $\lambda = \frac{1}{9}$ self-consistent densities show overestimation.
\item At the atomic sites the $F(N_e)$ and KS based densities are very close
and nonzero, whereas the $\lambda = \frac{1}{9}$ shows a disturbing feature
of almost zero density.
\item At the peaks on either side of the origin, the KS and $F(N_e)$
charge densities again are close, but the charge density by
$\lambda = \frac{1}{9}$ shows considerable overestimation.
\end{enumerate}

In Fig. 2 we have plotted the superposed free atom charge density
($0^{th}$ iteration density) represented by the curve b and the
self-consistent charge density for the functional
involving $\lambda = \frac{1}{9}$ by the curve c and $F(N_e)$ by
the curve a.
The self consistent charge density obtained using $\lambda = \frac{1}{9}$
shows improvement only at the origin. Contrary to this the
self consistent charge density using $F(N_e)$ shows a significant overall
improvement, both at the origin and at the peaks on either side
of the peak at the origin.
To get an idea of the nature of the forces obtained by the OFM and KS
dynamics, we have given the results for the vibrational frequencies for
Na, Mg and Li dimers in Table 2.
It is gratifying to note that the vibrational frequencies obtained by
OFM method are in very good agreement with the KS ones.

To assess the utility and performance of this method, it has been applied
to calculate the ground state geometries of a range of
small clusters. We report here our calculations on heteronuclear clusters
of $Li_nAl, n=1,8$ and a highly symmetric homonuclear cluster of $Al_{13}$
and clusters of $Na_n, n = 6,8$ using the $F(N_e)$ functional. The
results are compared with the ones reported by KS method.

The geometries of the heteronuclear $Li_nAl$ clusters are shown in Fig. 3
and the bondlengths and symmetries in Table III. along with the KS results.
Evidently, the present method not only reproduces the correct ground state
geometry with bondlengths within 5\% but also reproduces the two key
features observed in the more accurate KS  calculation \cite{cheng}.
\begin{enumerate}
\item The $Li_nAl$ clusters for $n < 3$ are two-dimensional whereas from
$n \geq 3$ the clusters become three-dimensional.
\item The Al atom gets trapped inside the Li atoms at $n = 6$.
\end{enumerate}
It can also be noted that as the number of atoms in the cluster increases,
the accuracy in the bondlengths appears to improve.
However, for the case of $Li_3Al$ and $Li_8Al$ we get the ground structure
configurations to be ideally symmetric rather than slightly Jahn Teller
distorted geometries of the KS calculation.

We have also investigated the $Al_{13}$ cluster since it shows an interesting
icosahedral geometry. The calculations were performed in two different ways.
First we started with a highly distorted icosahedron and applied the dynamical
quenching to get the equilibrium geometry.  In the second one, we started by
placing the Al atoms at the fcc lattice points and heated the cluster to
$300^{\circ} K$ and let the system span the configuration space for a few
thousand iterations. This was then followed by a slow cooling schedule.
It is very gratifying to note that in both the calculations the correct
icosahedron is obtained with a bondlength of 4.88 a.u. as compared to the KS
bondlength of 5.03 a.u. The error in the bondlengths being 3\%.
This strengthens our confidence in the ability of the method to reproduce
the correct ground state geometries with acceptable bondlengths.
In addition, we have also obtained the ground state geometries for Na$_6$,
Na$_8$ and Na$_{20}$ and have verified that the geometries obtained are
identical to those reported in \cite{urs}, with the bondlengths differing
by about 5\%.

\section {Conclusion}

In this work, we have presented the results obtained by using density based
{\it ab initio} MD for a variety of small clusters and demonstrated that
the method using approximate KE functionals is capable of yielding
bondlengths within an accuracy of 5\%.
Our calculations indicate that the ground state geometries and symmetries
of both homonuclear and heteronuclear clusters can be obtained within a
reasonable accuracy and timesteps of the order of 40 a.u. can be used
successfully for stable dynamics.
The $F(N_e)$ functional is shown to give considerable improvement over
standard $\lambda = \frac{1}{9}$ functional both in terms of charge
densities and bondlengths and is thus recommended.

We believe the method to be a promising tool in the  study of finite
temperature and dynamical properties of clusters.
So far, all the reported OFM calculations have been performed using
local pseudopotentials only and it would be interesting to
implement the nonlocal pseudopotentials and study the effect of
nonlocality on the bonding and binding properties of such clusters.
More work is required in this direction and the implementation of
nonlocality is under consideration.
It may be possible to expand the applications of OFM by incorporating
the nonlocal pseudopotentials and by employing more accurate KE
functionals.
It is hoped that the problems of current
interest in the field of clusters like fragmentation, dissociation,
interaction between clusters, which may involve large number of atoms
as well as more than one atomic species will be amenable by the present
technique.

\section {Acknowledgements}
Partial financial assistance from the Department of Science and Technology
(DST), Government of India and the Centre for Development of Advanced
Computing (C-DAC), Pune is gratefully acknowledged. Two of us (V. S
and D. N) acknowledge financial assistance from C-DAC. One of us (D. G. K.)
acknowledges P. Madden for a number of fruitful discussions on the OFM. We
also acknowledge K. Hermansson and L. Ojamoe for the MOVIEMOL animation
program.

\newpage

\vspace{3pt}

TABLE I. Comparison of the equilibrium bondlengths (in a.u) and binding
energies (in eV/atom) using the different kinetic energy functionals with the
KS self consistent method.

\vspace{0.3in}
\begin{tabbing}
\hspace{0.7in}\= hspace{0.5in}\= hspace{0.5in}\= hspace{0.5in}\=
hspace{0.5in}\= hspace{0.5in}\= hspace{0.5in}\= hspace{0.5in}\=\kill
 System \> \>    Bondlengths \> \> \> \> Binding energies   \\
\> $\lambda = \frac{1}{9}$ \> $\lambda = \frac{1}{5}$ \> $F(N_e)$ \>  KS \>
$\lambda = \frac{1}{9}$ \> $\lambda = \frac{1}{5}$ \> $F(N_e)$  \> KS \\ \\
$Na_2$ \> 5.67 \>  -   \> 5.69 \> 5.66$^a$ \> -0.116 \>    -   \> -0.867 \>
-0.71$^a$  \\
$Na_3$ \> 5.81 \> 5.99 \> 5.75 \> 6.00$^b$ \> -0.207 \> -0.281 \> -1.286 \>   -
       \\
$Mg_2$ \> 5.79 \>  -   \> 4.71 \> 6.33$^c$ \> -0.195 \>   -    \> -1.432 \>
-0.115$^c$ \\
$Mg_3$ \> 5.94 \> 5.81 \> 4.87 \> 5.93$^c$ \> -0.355 \> -0.526 \> -2.096 \>
-0.284$^c$ \\
$Al_2$ \> 5.74 \>  -   \> 4.14 \> 4.66$^d$ \> -0.261 \>   -    \> -1.389 \>
-1.06$^d$  \\
$Al_3$ \> 5.88 \> 5.57 \> 4.32 \> 4.74$^d$ \> -0.483 \> -0.733 \> -2.074 \>
-1.96$^d$  \\
$Li_2$ \> 5.87 \>  -   \> 5.51 \> 5.15$^e$ \> -0.102 \>   -    \> -0.891 \>   -
       \\
$Li_3$ \> 6.03 \> 6.11 \> 5.58 \> 5.3$^b$  \> -0.182 \> -0.256 \> -1.311 \>   -
       \\
$Si_2$ \> 5.35 \>  -   \> 3.74 \> 4.29$^f$ \> -0.371 \>   -    \> -0.56  \>
-0.6$^g$   \\
$Si_3$ \> 5.50 \>  -   \> 3.92 \> 4.10$^f$ \> -0.651 \>   -    \> -0.938 \>   -
       \\
\end{tabbing}
\noindent

$^a$Reference\cite{urs}    \hspace{0.5in} $^e$our own KS calculations

$^b$Reference\cite{martin} \hspace{0.5in} $^f$Reference\cite{feuston}

$^c$Reference\cite{kumar}  \hspace{0.5in} $^g$Reference\cite{tomanek}

$^d$Reference\cite{yang}   \hspace{0.5in}

\newpage

\vspace{3pt}

TABLE II. The vibrational frequencies (in $cm^{-1}$) of Na, Mg, Li dimer
using the OFM and KS self consistent method.

\vspace{0.3in}
\begin{tabbing}
\hspace{2in}\=\hspace{2in}\=\kill
 dimers \>  OFM  \> KS    \\
  Na    \> 167.4 \> 168   \\
  Mg    \> 107.3 \> 108.6 \\
  Li    \> 273.7 \> 311   \\
\end{tabbing}
\vspace{3pt}

TABLE III. The bondlengths between Li-AL of $Li_nAl, n=1,8$ using OFM
compared with those obtained by KS method \cite{cheng}. All the bondlengths
are in a.u.

\vspace{0.3in}
\begin{tabbing}
\hspace{1.5in}\=\hspace{1.5in}\=\hspace{1.5in}\=\hspace{1.5in}\=\kill
 system \>      OFM        \>      KS          \>\% error \> Symmetry       \\
$LiAl  $\>      4.77       \>      5.35        \>   10.8  \> $C_{\infty v}$ \\
$Li_2Al$\> 2 $\times$ 4.76 \> 2 $\times$ 5.22  \>    8.8  \> $C_{2v}$       \\
$Li_3Al$\> 3 $\times$ 4.79 \> 3 $\times$ 4.98  \>    3.8  \> $C_{3v}$       \\
$Li_4Al$\> 4 $\times$ 4.84 \> 2 $\times$ 4.82  \>    0.4  \> $C_{3v}$       \\
        \>                 \> 2 $\times$ 4.89  \>    1    \>                \\
$Li_5Al$\> 4 $\times$ 4.84 \> 4 $\times$ 4.74  \>    2    \> $C_{4v}$       \\
        \>      4.95       \>      5.13        \>    3.3  \>                \\
$Li_6Al$\> 6 $\times$ 4.79 \> 6 $\times$ 4.58  \>    4.5  \> $O_h$          \\
$Li_7Al$\>      4.97       \>      4.70        \>    5.7  \> $C_{1h}$       \\
        \> 2 $\times$ 4.92 \> 2 $\times$ 4.85  \>    1.4  \>                \\
        \> 2 $\times$ 4.88 \> 2 $\times$ 4.74  \>    2.9  \>                \\
        \> 2 $\times$ 4.89 \> 2 $\times$ 4.81  \>    1.6  \>                \\
$Li_8Al$\> 8 $\times$ 4.99 \> 8 $\times$ 4.82  \>    3.5  \> $D_{4d}$       \\
\end{tabbing}

\vspace{0.3in}
\newpage

\centerline{\bf Figure Captions}

\begin{enumerate}
\item The self-consistent charge densities of Al dimer.
Curve a  represents the $F(N_e)$ functional charge density, curve b represents
the $\lambda = \frac{1}{9}$ charge density and curve c denotes the charge
density obtained using the KS method.

\item Comparison of self-consistent charge densities by the
$F(N_e)$ (curve a) and $\lambda = \frac{1}{9}$ (curve c)
functional for Al dimer with the superposed ( $0^{th}$ iteration) free
Al atom charge density (curve b).

\item The ground state geometries of the $Li_nAl$ clusters for
$n = 1, 8$. The large sphere represents the Li atom and small sphere
represents the Al atom.
\end{enumerate}


\begin{references}
\bibitem[\ddag]{email}
Electronic Address: dinesh, vaishali, kanhere@unipune.ernet.in

\bibitem{car}
R. Car, M. Parrinello, Phys. Rev. Lett., {\bf 55}, 685(1985)

\bibitem{rem}
D. K. Remler and P. A. Madden, Molecular Physics, {\bf 70}, 921(1990)

\bibitem{footnote}
The algorithms which scale linearly with system size, based on density
matrix formulation have been proposed see:
W. Kohn, Chem. Phys. Lett. {\bf 208}, 167 (1993);
M. S. Daw, Phys. Rev. B. {\bf 47}, 10895 (1993);
X. P. Li, R. W. Nunes and D. Vanderbilt, Phys. Rev. B {\bf 47},
10891 (1993)

\bibitem{pear}
M. Pearson, E. Smargiassi and P.A. Madden, J. Phys. Condens.
Matter {\bf 5}, 3221 (1993)

\bibitem{shah}
V. Shah, D. Nehete and D. G. Kanhere, J. Phys: Condens.
Matter. {\bf 6}, 10773 (1994)

\bibitem{govind}
N. Govind, J. Wang and H. Guo, Phys. Rev. B. {\bf 50}, 11175 (1994)

\bibitem{smar}
E. Smargiassi and P. A. Madden, Phys. Rev. B. {\bf 51}, 117 (1995);
E. Smargiassi and P. A. Madden, Phys. Rev. B. {\bf 51}, 129 (1995)

\bibitem{niranjan}
N. Govind, J. L. Mozos and H. Guo, Phys. Rev. B. {\bf 51}, 7101 (1995)

\bibitem{smarg}
E. Smargiassi and P. A. Madden, Phys. Rev. B. {\bf 49}, 5220 (1994)

\bibitem{depristo}
A. E. DePristo and J. D. Kress, Phys. Rev. A. {\bf 35}, 438 (1987)

\bibitem{wang}
L. W. Wang and M. P. Teter, Phys. Rev. B. {\bf 45}, 13196 (1992)

\bibitem{hoh}
P. Hohenberg and W. Kohn, Phys. Rev. {\bf 136}, B864(1964)

\bibitem{ksh}
W. Kohn and L. J. Sham, Phys. Rev. {\bf 140}, A1133(1965)

\bibitem{bach}
G. B. Bachelet, D. R. Hamann and M. Schluter, Phys. Rev. B. {\bf 26},
4199 (1982)

\bibitem{parr}
R. G. Parr and W. Yang, {\it Density Functional Theory of Atoms
and Molecules} (O. U. P., Oxford, 1989)

\bibitem{acharya}
P. K. Acharya, L. J. Bartolotti, S. B. Sears and R. G. Parr, Natl. Acad.
Sci. U.S.A. {\bf 77}, 6978 (1980)

\bibitem{gaz}
J. L. Gazquez, and J. Robles, J. Chem. Phys. {\bf 76}, 1467 (1982)

\bibitem{ghosh}
S. K. Ghosh, L. C. Basbas, J. Chem. Phys. {\bf 83}, 5778 (1985)

\bibitem{press}
W. H. Press, B. P. Flannery, S. A. Teukolsky, W. T. Vetterling,
{\it Numerical Recipes}, Cambridge University Press, Cambridge (1987);
P. E. Gill, W. Murray, M. H. Wright, {\it Practical Optimization},
Academic Press, London (1988);
B. N. Pshenichny and Y. M. Danilin, {\it Numerical Methods in Extremal
Problems}, Mir publishers, Moscow (1978)

\bibitem{cheng}
H. P. Cheng, R. N. Barnett and U. Landman, Phys. Rev. B. {\bf 48},
1820 (1993)

\bibitem{urs}
U. Rothlisberger and W. Andreoni, J. Chem. Phys. {\bf 94},
8129 (1991)

\bibitem{martin}
J. L. Martins, R. Car and J. Buttet, J. Chem. Phys. {\bf 78},
5646 (1983)

\bibitem{kumar}
V. Kumar and R. Car, Phys. Rev. B. {\bf 44}, 8243 (1991)

\bibitem{yang}
S. H. Yang, D. A. Drabold, J. B. Adams and A. Sachdev, Phys. Rev. B
{\bf 47}, 1567 (1993)

\bibitem{feuston}
B. P. Feuston, R. K. Kalia, and P. Vashishta, Phys. Rev. B. {\bf 35},
 6222 (1987)

\bibitem{tomanek}
D. Tomanek and M. A. Schluter, Phys. Rev. B. {\bf 36}, 1208 (1987)

\end{references}
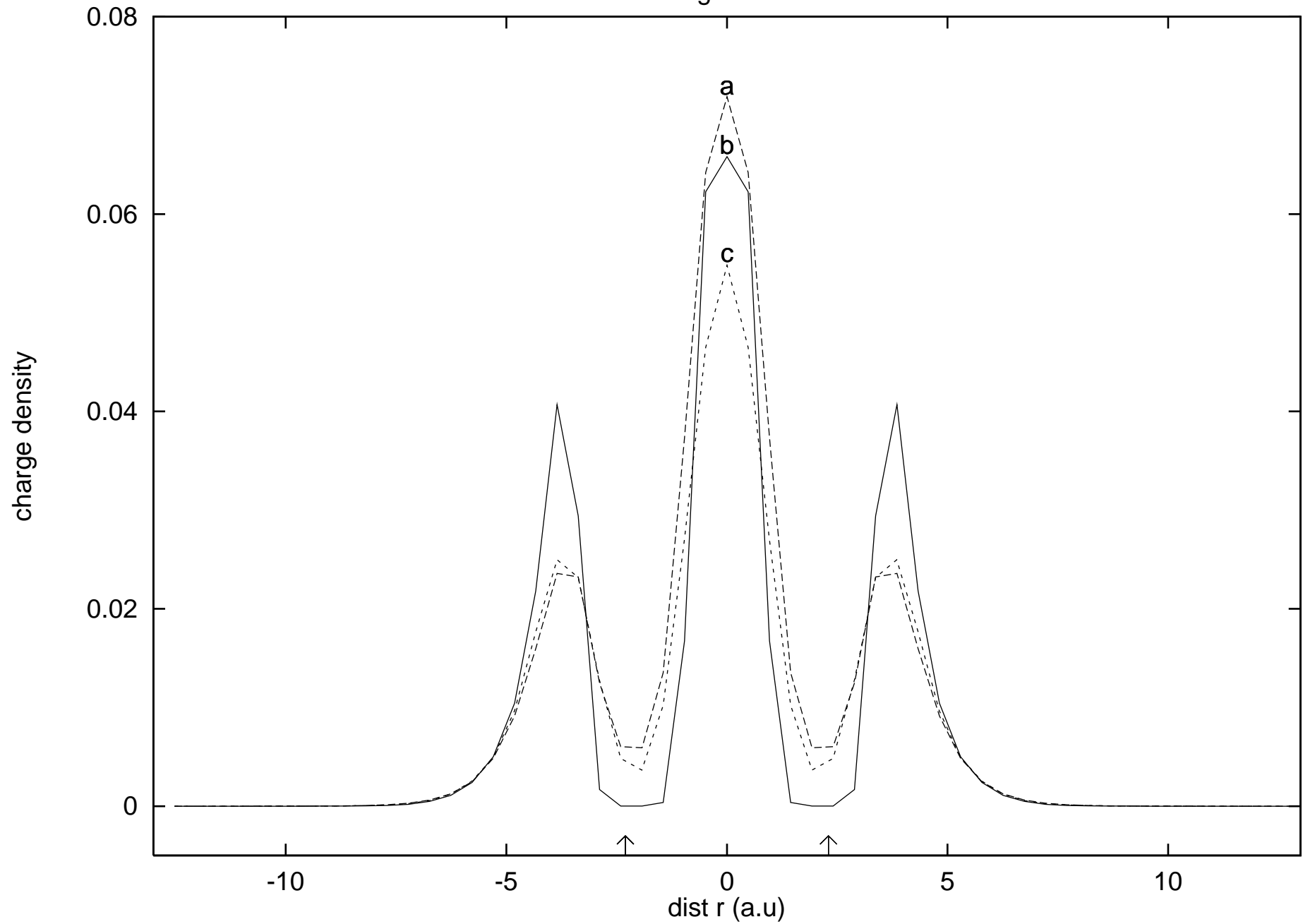
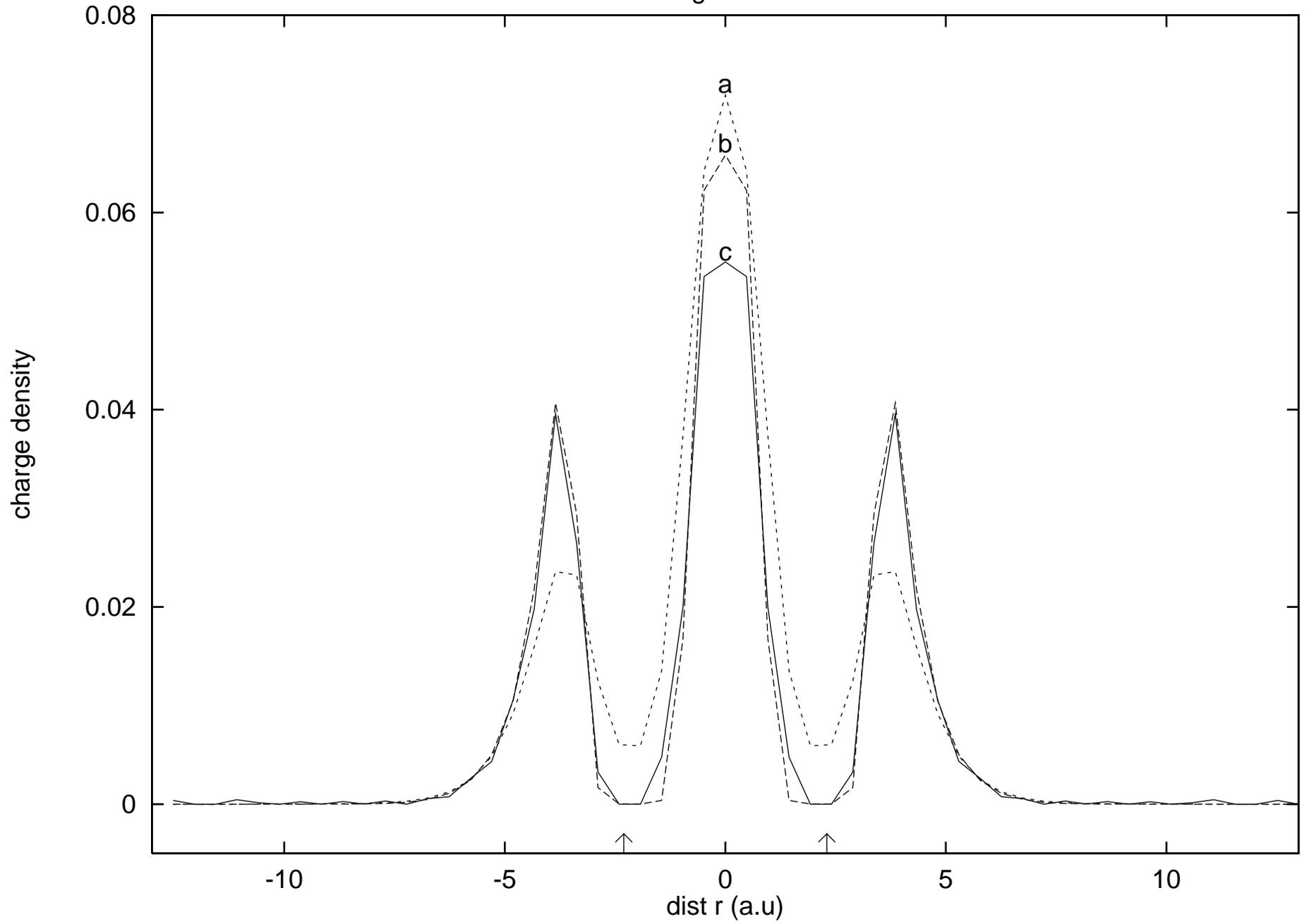
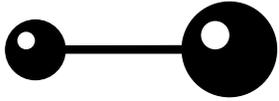
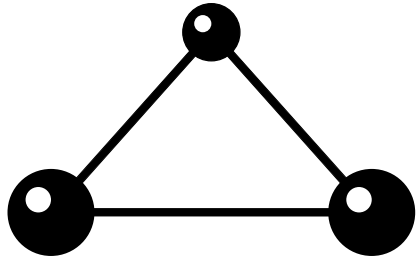
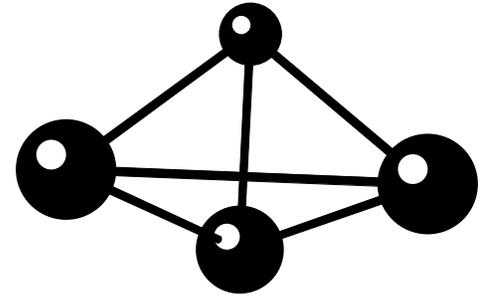
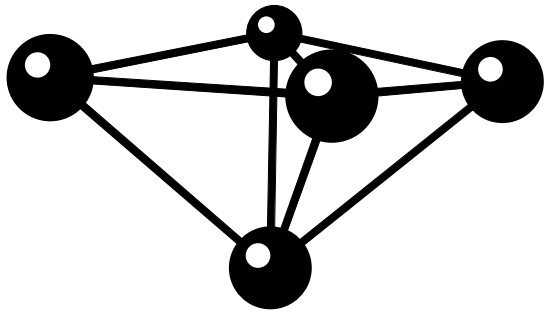
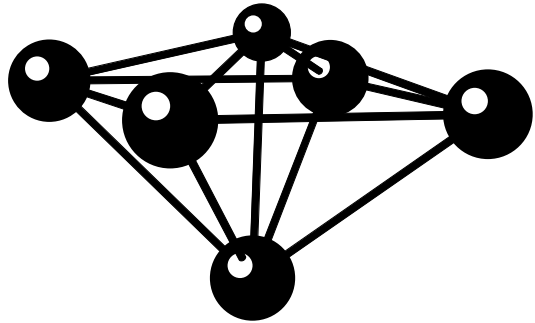
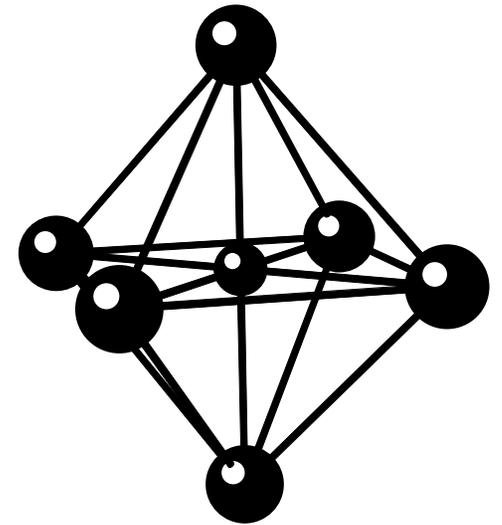
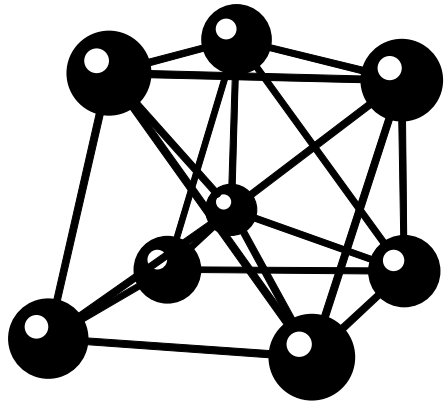
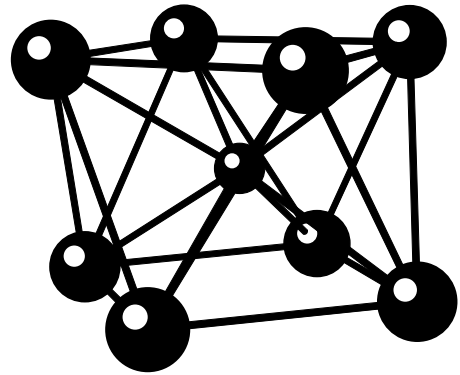
\end{document}